\documentclass[onecolumn]{emulateapj}
\slugcomment{To appear in ApJ 675}
\shorttitle{Induced formation of terrestrial planets}

\shortauthors{Thommes, Nagasawa \& Lin}

\begin{document}

\begin{abstract}
We revisit the ``dynamical shakeup" model of Solar System terrestrial planet formation, wherein the whole process is driven by the sweeping of Jupiter's secular resonance as the gas disk is removed.  Using a large number of 0.5 Gyr-long N-body simulations, we investigate the different outcomes produced by such a scenario.  We confirm that in contrast to existing models, secular resonance sweeping combined with tidal damping by the disk gas can reproduce the low eccentricities and inclinations, and high radial mass concentration,  of the Solar System terrestrial planets.  At the same time, this also drives the final assemblage of the planets on a timescale of several tens of millions of years, an order of magnitude faster than inferred from previous numerical simulations which neglected these effects, but possibly in better agreement with timescales inferred from cosmochemical data.  In addition, we find that significant delivery of water-rich material from the outer Asteroid Belt is a natural byproduct.  
\end{abstract}     

\title{Dynamical Shakeup of Planetary Systems II.  N-body simulations of Solar System terrestrial planet formation induced by secular resonance sweeping}

\author{E. W. Thommes}
\affil{Department of Physics and Astronomy, Northwestern University, Evanston, IL 60208, USA}

\author{M. Nagasawa}
\affil{Global Edge Institute, Tokyo Institute of Technology, 2-12-1,Ookayama, Meguro-ku, Tokyo 152-8550, Japan }
\author{D.N.C. Lin}
\affil{UCO/Lick Observatory, University of California, Santa Cruz,
CA 95064, USA}
\affil{and}
\affil{Kavli Institue of Astronomy and Astrophysics, Peking University, Beijing, China}

\email{thommes@northwestern.edu, nagasawa.m.ad@m.titech.ac.jp, lin@ucolick.org}


\section{Introduction}
\label{intro}
The formation of terrestrial planets from a disk of planetesimals proceeds through three different successive modes of accretion.  First runaway growth, wherein the most massive bodies have the shortest mass-doubling time (Wetherill \& Stewart 1989) produces a population of protoplanets within the planetesimals, which quickly detach themselves from the upper end of the planetesimal size distribution.  Growth stops being a runaway process once these bodies start to dynamically stir their surroundings 
\citep{1993Icar..106..210I}; at this point less massive protoplanets have a shorter mass-doubling time, so that there is a tendency for protoplanets to have (locally) similar masses.  This regime is usually called {\it oligarchic growth} 
\citep{1998Icar..131..171K}.  In the terrestrial region, the transition from runaway to oligarchic
growth already happens when the largest protoplanets are still many orders of magnitude below an Earth mass
\citep{2003Icar..161..431T}.  Thus, provided most collisions are not disruptive, most of the protoplanets' growth occurs in the oligarchic phase.  During this stage, closely separated protoplanets collide and merge with each other.  However, dynamical relaxation between widely separated protoplanets is suppressed by their tidal interaction with the residual disk gas 
\citep{2007ApJ...666..447Z}
 On the gas depletion time scale (3-10 Myr), they maintain 
a characteristic radial spacing of $\sim 10$ Hill radii ($r_H$), where $r_H = (M/3 M_*)^{1/3}r$ for a body of mass $M$ orbiting a primary of mass $M_*$ at a distance $r$.  The protoplanets' growth by sweep-up of planetesimals ceases when all planetesimals are gone, and the protoplanets have reached their {\it isolation masses}.  For the minimum-mass Solar nebula (MMSN) model (Hayashi 1981), the isolation mass ranges from $\sim 10^{-2}$ to $10^{-1}$ Earth masses (M$_\oplus$), i.e. lunar to martian mass.  
After the depletion of the disk gas, the full-grown oligarchs perturb each other onto crossing orbits and accrete each other in giant impacts, growing another decade in mass to produce bodies as large as Earth and Venus.  

N-body simulations have been the tool of choice for characterizing this last phase, since by the time all the planetesimals are locked up in tens to hundreds of protoplanets, $N$ is no longer an intractably large number.  In the past, such simulations included only gravitational forces, under the assumption that the final phase takes place when the nebular gas is long gone 
\citep{1998Icar..136..304C,1999Icar..142..219A,2001Icar..152..205C}.  Although individual simulation outcomes were highly stochastic, a number of characteristic features emerged:  Starting out with a roughly MMSN disk of protoplanets, (1) the final number and masses of bodies tended to be comparable to the Solar System; (2) the timescale for the the giant impact phase to play out was generally $\ga 10^8$ years; (3) the largest bodies formed ended up with eccentricities and inclinations significantly higher than those of Earth and Venus.  This last point has constituted a long-standing gap in our understanding of the formation of the terrestrial planets.  In the most general terms, solving this problem requires invoking an additional dissipative physical process.  Perhaps the most obvious candidate is leftover planetesimals, which can serve as a source of dynamical friction.  Though the timescale for oligarchic growth to finish in the terrestrial region is $\la 10^6$ 
(e.g. \citealt{2002ApJ...581..666K}), a reservoir of small bodies may be maintained by collisional replenishment 
\citep{2004ApJ...614..497G}.  As long as the nebular gas persists, it also acts as a source of damping for protoplanet eccentricities and inclinations via tidal interaction 
(e.g. \citealt{2000MNRAS.315..823P} and references therein).  Whatever the source of dissipation, the fundamental problem is that making the isolation-mass oligarchs' orbits cross and damping the eccentricities of the finished products are two conflicting objectives.  This conundrum is well-illustrated by the work of 
\cite{2002Icar..157...43K,2004Icar..167..231K}, who show that damping by the gas simply delays the onset of the giant impact phase.  At the time the gas is completely gone, the process is still incomplete; eccentricities are low but too many bodies remain.  

\cite{2006Icar..184...39O} performed N-body accretion simulations starting partway through the oligarchic growth process, with half the mass still in small planetesimals; due to the strong dynamical friction these exert on the larger bodies, the final planets attain eccentricities and inclinations comparable to Solar System values.  
However, 
this scenario is predicated on the assumption that a large population of small planetesimals is present during the final growth phase of the protoplanets; it is questionable whether this is realistic.
Simulations incorporating detailed modelling of collision outcomes, including both coagulation and fragmentation 
\citep{2005ApJ...625..427L}
show that the population of residual planetesimals  is small compared with that of the growing protoplanets.  However, this is not a unanimous conclusion; the simulations of \cite{2004ApJ...613L.157A} produce a lower accretion efficiency in high-velocity collisions, suggesting that a significant steady-state population of collisional debris could in fact be maintained late into the formation process.
Observationally, collisions among a long-lived population of small planetesimals would lead to the continual generation of dust grains on the growth timescale of the protoplanets. The protracted presence of such a rich supply of dust grains would appear to be inconsistent with the rapid (a few Myr) depletion of NIR, MIR, and FIR signatures of grains throughout the disk (\citealt{2007prpl.conf..573M} and references therein).  The simulations presented by \cite{2006Icar..184...39O} also neglect the presence of any residual gas in the disk.  In a gas-free environment, there is no artificial transition between the oligarchic growth and  giant impact phases. Dynamical isolation of oligarchs is associated with low-eccentricity orbits 
\citep{2007ApJ...666..423Z}.
At $\sim 0.5-3$ AU in the MMSN, the time scale for the embryos to attain isolation mass 
  ($\sim$ a few $10^5$ yr) is generally much shorter than the disk depletion time observed
  in protostellar disks 
\citep{2001ApJ...553L.153H,2001ASPC..244..245M}   
   The results of \cite{2002Icar..157...43K,2004Icar..167..231K} cited above suggest that in reality orbit crossing is suppressed until the gas is nearly gone, effectively stalling growth at the end of oligarchy.  In that case, the giant impact phase begins with almost all the mass in protoplanets, and little mass in planetesimals to provide dynamical friction.

\cite{2005ApJ...635..578N,2007prpl.conf..639N} developed a model which addresses this ``Catch-22", demonstrating how the excitation of protoplanets onto crossing orbits and subsequent damping of eccentricities could have coexisted in the early Solar System as near-simultaneous processes.  Their scenario invokes a dissipating gas disk but also includes its effect on the pericenter precession of the embedded (proto)planets.  The net precession of a given protoplanet is determined by the sum of the contributions from Jupiter, Saturn and the disk.  When this precession rate matches that of one of the giant planets, a precession (or secular) resonance between the two bodies occurs, resulting in rapid pumping of the protoplanet's eccentricity.  As the gas disk is depleted, the location of the inner secular resonance with Jupiter (denoted $\nu_5$) sweeps inward, transiting the asteroid belt and then the terrestrial region.  One after another, the protoplanets are hit by the resonance, resulting in an inward-passing wave of orbit-crossing.  As shown by 
\cite{2005ApJ...635..578N}, this tends to result in numerous collisions between protoplanets.  Thus, with the passage of the $\nu_5$ resonance serving as a trigger, the final stage of terrestrial planet formation is initiated {\it as} the disk dissipates, rather than afterwards.  If the whole process is finished while sufficient gas is left, the end products are then damped to low eccentricities and inclinations.  This ``dynamical shakeup" model therefore provides a pathway to producing terrestrial planets with the nearly circular and coplanar orbits of Earth and Venus.  Since the formation timescale is tied to the dissipation time of the gas, everything happens in of order $10^7$ years, an order of magnitude faster than in gas-free  N-body simulations of the giant impact phase.   

In the final stage of their assemblage, the planets' excited eccentricities enables them to cross each other's orbits, but by itself this would also suppress the gravitational focusing effect.  In the limit that the orientation of the planets' orbits is totally randomized, their growth timescale would then be their dynamical timescale divided by their area filling factor 
\citep{2004ARA&A..42..549G}
which, at 0.5-3 AU  in a MMSN, is $\sim 10^8$ yr.  However, the combination 
  of eccentricity excitation due the sweeping secular resonance with damping due to planet-disk tidal interaction tends to align the protoplanets' longitudes of periapse. This effect 
  reduces the protoplanets' relative velocities while still allowing their orbits to cross.  
  Consequently, gravitational focusing continues to be effective, permitting growth on a timescale comparable to that of the resonance-sweeping \citep{2005ApJ...635..578N}.

Here, we perform N-body simulations to assess the range of outcomes generated by the dynamical shakeup model.  In particular, we focus on the issue of how readily analogs of our Solar System's terrestrial region are produced.  As an initial condition, we begin with an educated guess for the state of the Solar System at the time oligarchic growth has concluded interior to the orbit of Jupiter.  Though the individual simulations are too computationally expensive to allow an exhaustive parameter study,  we vary a number of key parameters (gas disk height profile, disk dispersal time, giant planet eccentricity) to get an idea of how sensitive the model is to them.   
In \S \ref{background}, we give a brief summary of the key features of the dynamical shakeup model.  In \S \ref{simulation setup}, we describe the setup of the numerical simulations.
The outcome of our baseline set of simulations is described in \S \ref{baseline_case}, as well as the criteria by which we assess whether or not a given result is ``Solar System-like".  In \S \ref{changing t_e}-\ref{set C}, we describe the effect of varying a number of the model parameters.  The issue of water delivery is considered in \S \ref{water}.  The results are discussed in \S \ref{discussion}.

\section{The dynamical shakeup model}
\label{background}

We begin with a brief summary of the terrestrial planet formation model of 
\cite{2005ApJ...635..578N}.  The key to this model is that the dissipating gas disk interacts with the embedded protoplanets in two different ways:  It damps their eccentricities, and it changes their apsidal precession rate.  The former effect is due to tidal torques, and operates on a timescale
\begin{equation}
\tau_{\rm e,tide} \equiv -\frac{e}{\dot{e}} \simeq \left( {M_\ast
\over M_p} \right) \left( {M_\ast \over \Sigma_g a^2} \right)
\left( {H \over a} \right)^4 \Omega_k ^{-1} \label{t_e_tide}
\label{tidal damping rate}
\end{equation}
where $M_p$ is the body's mass, $a$ its semimajor axis, and
$\Sigma_g$, $H$ and $\Omega_k$ the gas surface density, disk scale
height and Keplerian angular velocity, respectively, at $r=a$
\citep{1989ApJ...345L..99W,1993Icar..106..274W,1993ApJ...419..166A}.
Tidal interaction,
specifically the mismatch between inner and outer Lindblad
torques, can also remove angular momentum $J$ from the bodies' orbits
on a timescale $\tau_{\rm J,tide} \equiv -J/\dot{J}$, in what is commonly called type I migration.
Two-dimensional calculations in an unperturbed thin disk produce
$\tau_{\rm J,tide}\sim 10^2 \tau_{\rm e,tide}$ 
\citep{1993ApJ...419..166A,1997Icar..126..261W,2000MNRAS.315..823P}.  For the regime we are considering---0.1 to 1 M$_\oplus$ objects embedded in a gas disk significantly depleted below its MMSN surface density---the latter effect is relatively unimportant and for simplicity we neglect it.  However, even with $\dot{J}=0$, eccentricity damping alone will
cause a protoplanet to migrate inward by removing energy from its
orbit.  Setting
\begin{equation}
\dot{J} =\sqrt{G M_*} M_p \left [\frac{\dot{a}\sqrt{1-e^2}}{2 \sqrt{a}} -
\frac{\sqrt{a} e \dot{e}}{\sqrt{1-e^2}} \right ]=0,
\end{equation}
we obtain an expression for the migration rate in terms of the
eccentricity-damping timescale:
\begin{equation}
\dot{a} = a\frac{e^2}{(1-e^2)}\frac{2}{t_{e,{\rm tide}}}
\label{migration timescale}
\end{equation}

We now consider what happens when this protoplanet are also subject
to secular perturbations.  The perturbations from the giant planets plus the gas disk induce a net precession 
\begin{equation} 
g_{\rm p}=g_{\rm p,J}+g_{\rm p,S}+g_{\rm p,disk}.
\end{equation}
Interaction with a planet changes the protoplanet's eccentricity.  The eccentricity rate of change due to Jupiter is
\begin{equation}
\dot{e}_{\rm p,J}=-\frac{|C_{\rm p,J}|}{g_{\rm p,J}} \sin{\eta}
\end{equation}
where $\eta \equiv (\varpi_p-\varpi_J)$ is the angle between the pericenters of the orbits, $C_{\rm p,J}\equiv -b^{(2)}_{3/2}(\alpha)/b^{(1)}_{3/2}(\alpha)$, $b_l^{k}$ are the Laplace coefficients, and $\alpha = a_{\rm p}/a_{\rm J}$.  Thus, variations in $e_{\rm p}$ are sinusoidal in general.  However, when the protoplanet approaches the $\nu_5$ secular resonance with Jupiter, $g_{\rm p} \approx g_{\rm J}$, so that $\eta$ varies slowly and, for $\pi < \eta < 2 \pi$, the protoplanet's eccentricity grows monotonically.  Jupiter itself precesses due to its interaction with Saturn as well as the disk (neglecting the effect of the protoplanets): $g_{\rm J}=g_{\rm J,S}+g_{\rm J,disk}$.  Thus, the condition for the $\nu_5$ secular resonance is
\begin{equation}
g_{\rm p,J}+g_{\rm p,S}+g_{\rm p,disk} = g_{\rm J,S}+g_{\rm J,disk}
\end{equation}
Similarly, the secular resonance with Saturn, called $\nu_6$,
occurs when
\begin{equation}
g_{\rm p,J}+g_{\rm p,S}+g_{\rm p,disk} = g_{\rm S,J}+g_{\rm S,disk}.
\end{equation}
The secular perturbations from Jupiter and Saturn modulate the
protoplanet's eccentricity without themselves modifying its
semimajor axis. In other words, secular interactions do the
opposite of the eccentricity damping prescription we assume above:
they change only angular momentum, not energy, of the
protoplanet's orbit.  If the disk is simultaneously exerting tidal
damping on the body on a timescale $\tau_{e,{\rm tide}}$, the net
eccentricity rate of change is
\begin{equation}
\dot{e}_{\rm p} = \dot{e}_{\rm p,J}+\dot{e}_{\rm p,S}-
\frac{e}{\tau_{e,{\rm tide}}} \label{net ecc rate of change}.
\end{equation}
As long as a nonzero eccentricity is
maintained, the body continues to migrate inward as per Eq.
\ref{migration timescale}.

As discussed in 
\cite{2005ApJ...635..578N}, under the right conditions the combination of resonant eccentricity excitation with the orbital decay induced by gas disk eccentricity damping (Eq. \ref{migration timescale}) will cause a body to temporarily keep pace with an inward-sweeping secular resonance.  They show that the minimum semimajor axis to which the $\nu_5$ resonance can drive a body is roughly given by 
\begin{equation}
a \sim 2 \left ( \frac{\tau_n}{\rm Myr} \right )^{-1/4} \left (\frac{M}{M_\oplus} \right )^{-1/4} \left (\frac{e_J}{0.05} \right )^{-1/2} {\rm AU}
\label{secular resonance condition}
\end{equation}
where $\tau_n$ is the gas nebula depletion timescale.

Migration of bodies ahead of a secular resonance is somewhat
analogous to capture and migration in mean-motion resonances,
however there are some important distinctions.  First, the
captured body exchanges only angular momentum with its captor; to
exchange energy and actually migrate requires an additional
interaction, in our case tidal damping by the disk. Secondly,
since the captured body always loses energy to the disk, only
inward migration is possible; this means that in a depleting disk
the outward-sweeping secular resonances (e.g. the $\nu_5$ at
$a>a_J$) are {\it not} able to entrain bodies.  Most importantly,
however, secular resonances have a much farther reach than
mean-motion resonances.  As an illustration, the present-day
location of the $\nu_5$, 0.6 AU, corresponds to a 25:1 mean-motion
resonance with Jupiter; being of 24th order, this resonance is
extremely weak and negligible for most practical purposes.

\section{Numerical simulation setup}
\label{simulation setup}
We perform simulations using code based on the SyMBA symplectic
integrator 
\citep{1998AJ....116.2067D}.  SyMBA is fast for
near-Keplerian systems; it requires a minimum of only about twenty timesteps per
shortest orbit, while undergoing no secular growth in energy
error.  To investigate the effects of a dissipating gas disk, we
add two components to the base SyMBA code:  A dissipational force
to model the effect of resonant planet-disk interactions, and a
change to the central force to model the effect of the disk's
gravitational potential on the precession rates of the embedded
bodies.  Following \cite{2000MNRAS.315..823P}, the former is modeled as
\begin{equation}
\vec{a}_{\rm damp} = -2 \frac{\left ( \vec{v} \cdot \vec{r}\right ) \vec{r}}{r^2
\tau_{e,{\rm tide}}} \label{edamp_acc}
\end{equation} 
where $\tau_{e,{\rm tide}}$ is given by Eq. \ref{t_e_tide}.  Since
this is a purely radial force, it changes only the energy, not the
angular momentum, of the body's orbit.

In applying the gravitational potential of the gas disk, we assume
that it has a $\Sigma \propto r^{-1}$ profile. This allows us to
use a simple Mestel potential \citep{1963MNRAS.126..553M} for the disk:
\begin{equation}
\Phi(r)=v_c^2 \ln r, \label{mestel}
\end{equation}
where
\begin{equation}
v_c = \sqrt{2 \pi G \Sigma_0 r_0}
\end{equation}
and $\Sigma_0$ is the surface density at radius $r_0$.
This disk profile is shallower than the MMSN, which has $\Sigma
\propto r^{-3/2}$.  
However, it agrees with that inferred for steady-state accretion disks based the standard $\alpha$ presciption 
\citep{1973A&A....24..337S}
and the equilibrium 
temperature profile for irradiated disks (see Eq. \ref{eq:diskheight}) 
\citep{1981PThPh..70...35H}.
In any case, the precession induced by
an unperturbed disk in an embedded body depends most strongly on
the material near the body's orbit 
\citep{1981Icar...47..234W}.  Furthermore, the
precession of Jupiter and Saturn, which for plausible nebula
parameters open at least a partial gap in the disk, is dominated by their mutual
interaction, with the disk being of secondary importance. 
Thus,
the sweeping of the secular resonances is primarily a consquence
of the changing potential felt by the protoplanets, rather than by the
giant planets.  
\citep{2005ApJ...635..578N}.  This means that a given location in the disk will always be swept by a secular resonance when the surface density at that location drops to a given approximate value, without strong
dependence on the distribution of gas further away. Consequently,
the ratio of secular eccentricity excitation to eccentricity
damping---the key quantity in our model---is not strongly
dependent on the assumed profile.  For simplicity, we do not apply the disk potential to the giant planets.

\section{Putting it all together:  Simulations of secular-resonance-induced terrestrial
planet accretion}
\subsection{The baseline case}
\label{baseline_case}
We now construct a full analogue of the inner Solar System
including the giant planets, starting at a stage where isolation
has been reached everywhere.  The terrestrial/asteroid belt
region, from 0.5 to 3 AU, is populated with protoplanets.  We use
an overall surface density of solids
\begin{equation}
\Sigma_s = 7 \left ( \frac{r}{\rm 1\,AU}\right )^{-1} {\rm
g\,cm^{-2}};
\end{equation}
this is divided up into bodies spaced 10 Hill radii apart, in keeping with outcomes of simulations of oligarchic growth.  This results in
bodies ranging in mass from several times $10^{-2}\,M_\oplus$ at
0.5 AU, to several times $10^{-1}\,M_\oplus$ at 3 AU. Initial
protoplanet eccentricities and inclinations (in radians) are $\la
10^{-3}$.  

A natural starting point for our simulations is the first appearance of the giant planets' secular efffects---that is, the time when the giant planets themselves first appear on the scene.  We assume the giant planets form via core accretion, so that the majority of their mass gain happens in a rapid final growth spurt, on a timescale of $< 10^5$ years, as they accrete their gas envelopes \citep{1996Icar..124...62P}.  Thus, we begin with a proto-Jupiter and proto-Saturn, each near its ``crossover" state just before the onset of runaway gas accretion, when the solid core and the (still hydrostatic) gas envelope have similar masses.  In keeping with the models of \cite{1996Icar..124...62P}, we choose the proto-giants' masses to be 30 $M_\oplus$.  Shortly after the start of the simulation, at $1.5
\times 10^5$ yrs and $5 \times 10^5$ yrs respectively, each
planet's mass is linearly increased on a timescale of $10^5$ years, to its current mass (310
$M_\oplus$ and 95 $M_\oplus$, respectively).   The growth onset times are largely arbitrary, simply chosen so that (i) Jupiter grows first, (ii) there is no strong coincidence in growth, i.e. the difference between is longer than the growth timescale, while (iii) the time for both planets to appear is still short compared to the gas disk dissipation timescale.  In general, planets in a gas disk are thought to be subject to significant migration; theory (see \citealt{2007prpl.conf..655P}
 for a review) and the presence of extrasolar hot Jupiters argue for this.  On the other hand, the orbital radii of Jupiter and Saturn are consistent with where theory predicts their birthplace \citep{2003Icar..161..431T}, thus the Solar System may be a case where little migration took place.  At the same time, in addition to planet-disk interaction, scattering of planetesimals among Jupiter, Saturn, Uranus and Neptune is thought to have produced some additional migration by moving the planets' orbits apart \cite{1984Icar...58..109F}.  A more closely-spaced Jupiter and Saturn moves the $\nu_5$ resonance further from the Sun, thus insofar as the planets' migration occurred during the dispersal of the gas disk, the inward sweep of the $\nu_5$ resonance would have been due to superposition of gas dispersal and Jupiter and Saturn moving apart.  Given the uncertainly over how much migration took place, we opt for the simplest approach and place the proto-giants at their present-day semimajor axes, 5.2 and 9.5 AU.  Caveats related to the issue of planet migration are discussed in \S \ref{discussion}.
 
 At their initial masses proto-Jupiter and -Saturn are too small to open gaps in the gas disk and thus ought to be subject to the strong tidal eccentricity damping of Eq. \ref{tidal damping rate}.  However, once a planet has grown to gap-opening mass, the eccentricity evolution is less clear.  One possibility is that the saturation of eccentricity-reducing corotation, which compete with eccentricity-exciting Lindblad resonances, causes a net pumping of a gap-opening planet's eccentricity \cite{2003ApJ...585.1024G}; some recent simulations have indeed shown such behavior, with gap-opening planets reaching eccentricities in excess of 0.1 \citep{2006ApJ...652.1698D}.  Just prior to gap formation quenching the proto-gas giants' growth, the characteristic time scale for their progenitors to double their mass is $\sim 10^{2-3}$ yr.  These rapid changes destabilize the orbits of nearby embryos and lead to close encounters 
\citep{2007ApJ...666..447Z}
Direct collisions results in merger events which contribute to the diversity of gas giants' internal structure and metallicity, while close scatterings can lead to ejection of residual embryos, but also excite the gas giants' eccentricities. From the perspective of our model, higher giant planets are actually preferable because they increase the strength of the secular resonance \citep{2005ApJ...635..578N}.  Thus we choose initial eccentricities of 0.075 for the giant planets in our baseline model, 1.5 times the current value of $\sim 0.05$, though we also investigate the effect of lowering eccentricities to their current values in \S \ref{set C} 
 
The gas component of the system is modelled as a disk with
a power-law surface density which exponentially decays in time:
\begin{equation}
\Sigma_g = 2000 \left ( \frac{r}{\rm 1\,AU}\right )^{-1} \exp
\left ( \frac{-t}{\tau_n} \right ) {\rm g\,cm^{-2}}
\label{gas disk Sigma}
\end{equation}
We begin here with a nebula dispersal timescale of $\tau_n=5 \times 10^6$ years.  For the purpose of computing the tidal eccentricity damping rate,
$\tau_{\rm e,tide}$, we assume a flared disk scale height like
that of 
\cite{1981PThPh..70...35H}:
\begin{equation}
H = 0.05 \left ( \frac{r}{\rm 1\,AU}\right )^{5/4}{\rm\,AU}.
\label{eq:diskheight}
\end{equation}

\begin{figure}
\plotone{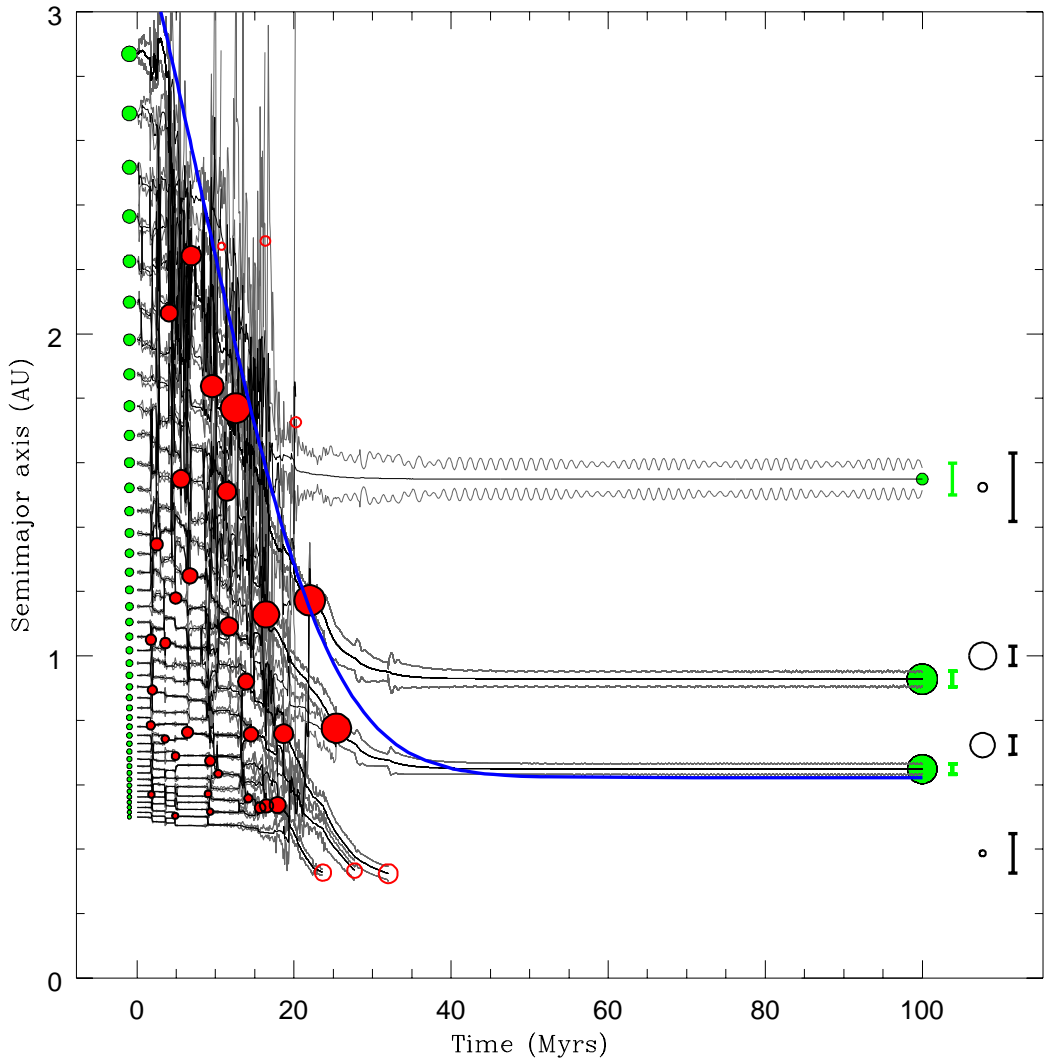}
\caption{The evolution of Run A1-9 (first 100 Myrs).  Initial protoplanet semimajor
axes and masses are shown at left (solid green circles, area
$\propto$ mass). Each body's semimajor axis (black) as well as
peri- and apocenter distance (gray) are then plotted as a
function of time, together with any mergers that occur (solid red
circle, area $\propto$ merger product), and bodies lost either by ejection,  or by removal at the inner simulation domain boundary of 0.3 AU (open red circles).  The path of the $\nu_5$
resonance is also shown (blue).  Final semimajor axes and masses
(solid green circles, area $\propto$ mass) are shown on the right
side, with the vertical error bars indicating the final peri- and
apocenter locations of each.  For comparison, the present-day
Solar System is shown at far right (black open circles, error
bars).} \label{apa_with_snapshots_9_PAPER}
\end{figure}

\begin{figure}
\plotone{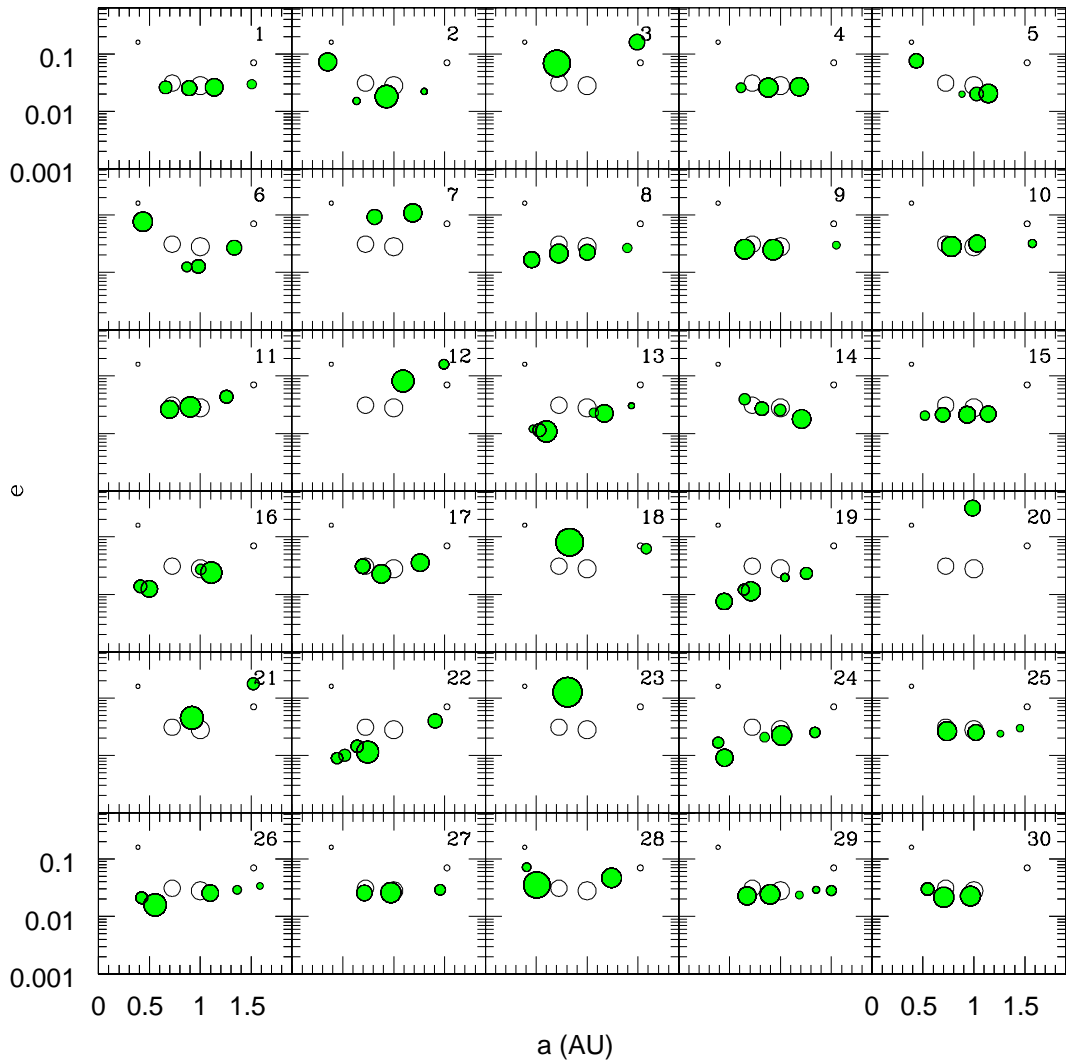}
\caption{All 30 runs in Set A1 after $10^8$ years.  Semimajor axis
and eccentricity of each surviving body are plotted (solid green
circles, area $\propto$ mass), with the current values of Mercury,
Venus, Earth and Mars (empty black circles) shown for comparison.}
\label{baseline_all}
\end{figure}

\begin{figure}
\plotone{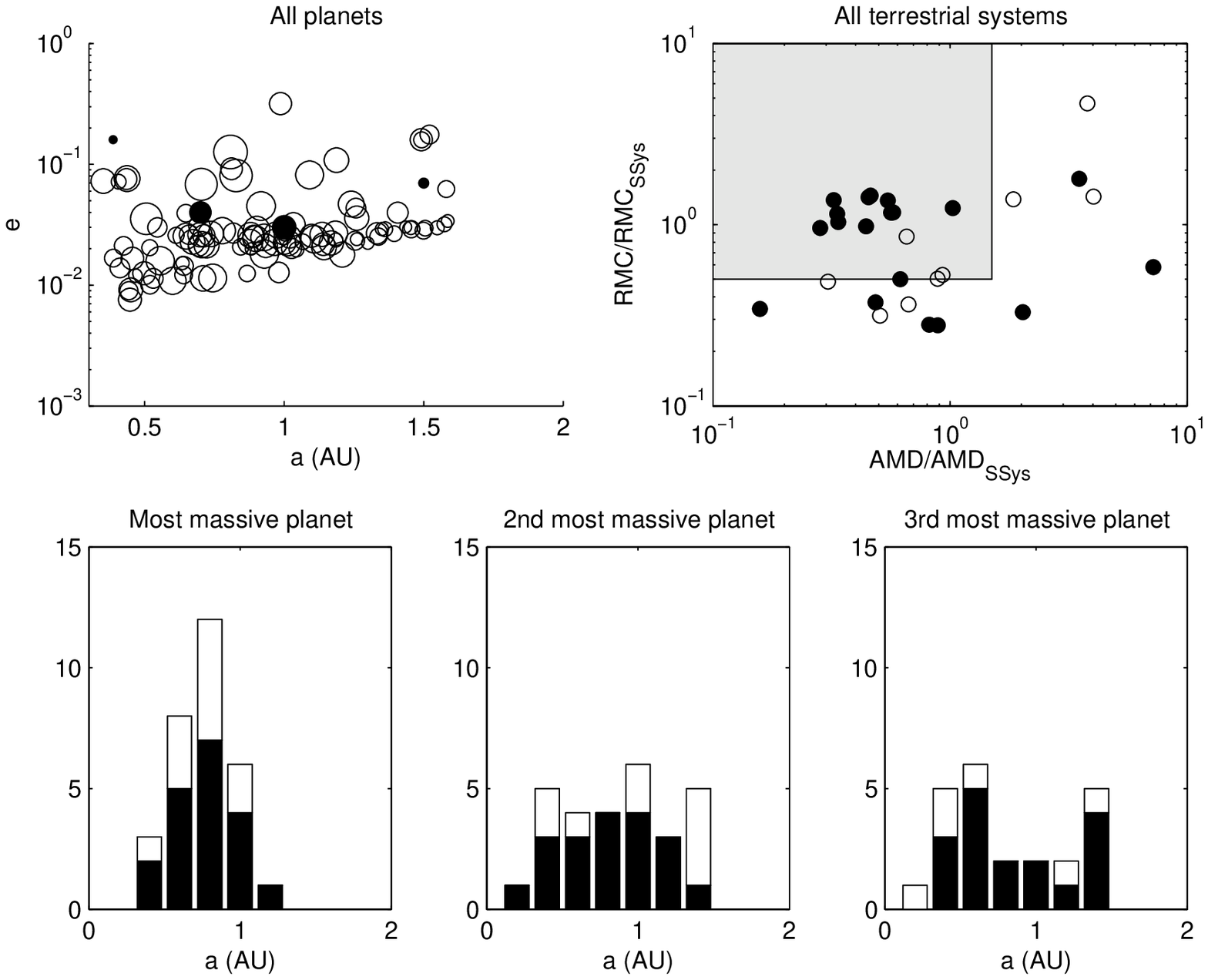} 
\caption{All 30 runs in Set A1.  Top left: semimajor axis and eccentricity of all planets (empty circles), with the Solar System terrestrial planets shown for comparison (solid circles).  Top right:  Angular momentum deficit (AMD) and radial mass concentration (RMC) of all individual runs.  Solid circles denote runs left with at least two planets of mass $\geq 0.5$ M$_\oplus$, but not more than four planets in total.  Bottom three panels:  final radial distribution of the first, second and third most massive planet.  The solid parts of the histograms show the subset of systems denoted by solid circles in the top right panel.}
\label{Set_A1_multipanel}
\end{figure}

We perform a set of 30 such simulations (hereafter Set A1), with
initial conditions identical except for a random variation in the
initial orbital phases of the protoplanets.  We use a base timestep of 0.015 years, and an inner edge at 0.3 AU, thus providing a resolution of just over ten steps per smallest orbit.  The very inner region of of our simulation domain is thus somewhat marginally resolved in time, but we deem this an acceptable compromise to keep the computational cost reasonable.  The inner boundary of the actual sweeping process is the final location of the $\nu_5$ resonance, between 0.6 and 0.7 AU; the resolution here is a comfortable 30 steps per orbit.

Each is run to $5 \times 10^8$ years of simulated time, which takes several weeks of wall time on a CPU of CITA's McKenzie cluster.  One example is shown
in Fig. \ref{apa_with_snapshots_9_PAPER}.  Clearly visible is the
way in which the inward-sweeping $\nu_5$ resonance shepherds the
protoplanets ahead of itself, bringing about orbit crossing and
numerous mergers.  A Mars-mass body is left behind by the $\nu_5$
early, after less than 20 Myrs of simulation time, with a
semimajor close to that of Mars.  An $\sim$ Earth-mass body leaves
the resonance at a bit less than 1 AU, while a second body of
similar mass follows the resonance almost all the way to its final
location between 0.6 and 0.7 AU.  Both bodies end up with averaged
eccentricities which are actually slightly {\it lower} than those
of Earth and Venus.  Thus, a good approximation of the present-day
inner Solar System is formed in 20-30 Myrs.  Around this time,
also, ``Earth'' and ``Venus'' suffer their last impact with another
protoplanet.

Fig. \ref{baseline_all} shows the outcomes of all 30 Set A1
simulations.  All have been run to $5 \times 10^8$ years, and have thus reached long-term stable configurations.  A1.4, A1.9 (the
case shown in Fig. \ref{apa_with_snapshots_9_PAPER}), A1.10 and
A1.27 are particularly close analogs to the inner Solar System:  The
eccentricities of the largest bodies are only a few times $10^{-2}$, and the masses and orbital spacings are comparable to those of the
terrestrial planets.  At the other extreme, there are number of systems which look decidedly different from the Solar System.  These can be grouped into two categories:  Systems where the largest planets have attained high eccentricities (e.g. A1.7, A1.23), and systems in which, although the planets have low eccentricities, they are more numerous and crowded than in our own terrestrial region (e.g. A1.13, A1.19).  

Fig. \ref{Set_A1_multipanel} shows a set of plots which summarize the outcomes of all the Set A1 runs in a compact form.  First, it shows (top left) a combined plot of all planets in all 30 systems.  From this, we see that the majority of planets have eccentricities of a few times $10^{-2}$, comparable to the Solar System.  There is a tendency for smaller planets to have lower eccentricities; these are predominantly the members of the group of more crowded systems (i.e. having undergone fewer mergers) mentioned above.  

The figure also shows (top right) a plot of the radial mass concentration versus the angular momentum deficit of each system.  Past simulations of late-stage terrestrial planet formation have systematically failed to reproduce values similar to the Solar System for these two key quantities 
\citep{2001Icar..152..205C}, thus we focus on them in quantifying whether or not a given outcome resembles the Solar System.  The radial mass concentration is a measure of the degree to which mass is radially localized, used by 
\cite{2001Icar..152..205C}:
\begin{equation}
RMC=max \left ( \frac{\displaystyle \mathop{\sum}_{j} M_j}{\displaystyle \mathop{\sum}_{j} M_j \left [ \log_{10}(r/a_j)\right ]^2} \right ) 
\end{equation}
The quantity in brackets measures the crowdedness of the system at a given value of $r$; the value we compute for $RMC$ is the maximum value this quantity takes on throughout the final terrestrial planet system.

The angular momentum deficit is a measure of the system's overall deviation from circular, coplanar orbits.  In normalized form, it is given by:
\begin{equation}
AMD=\frac{\displaystyle \mathop{\sum}_{j} M_j \sqrt{a_j}\left [1 - \sqrt{1-e_j^2}\cos i_j \right ]}{\displaystyle \mathop{\sum}_j M_j \sqrt{a_j}}
\label{AMD}
\end{equation}
\citep{1997A&A...317L..75L}.
Plotted values are normalized to the values for the Solar System's terrestrial planets.  We choose  $RMC > 0.5 RMC_{\rm Solar\,System}$ and $AMD < 2 AMD_{\rm Solar\,System}$ as conditions for a terrestrial planet system to be ``Solar System-like".  
Additionally, we impose the conditions that the system contain at least two planets with mass $> 0.5$ M$_\oplus$, but not more than four planets in total.  Systems which satisfy these two latter conditions are plotted as filled circles.  Fig. \ref{Set_A1_multipanel} shows that a total of eleven system satisfy all conditions, thus about a third of all simulations in the set result in a credible Solar System analog.  

Fig. \ref{Set_A1_multipanel} also shows the overall radial distribution of the most massive, second most massive and third most massive planets in Set A1.  For the two most massive bodies, the distribution is peaked near 1 AU, while for the third most massive planet, the distribution avoids this region.  

\subsection{Changing the strength of tidal damping}
\label{changing t_e}

To investigate the parameter dependence of our model, we begin by varying the strength of the tidal
damping.  We parameterize this by changing the the disk scale
height; recall that $\tau_{\rm e, tide} \propto H^{4}$ (Eq.
\ref{t_e_tide}).  Since the disk is approximated as
two-dimensional for the purpose of computing its gravitational
potential (Eq. \ref{mestel}), the potential remains unchanged.

\begin{figure}
\plotone{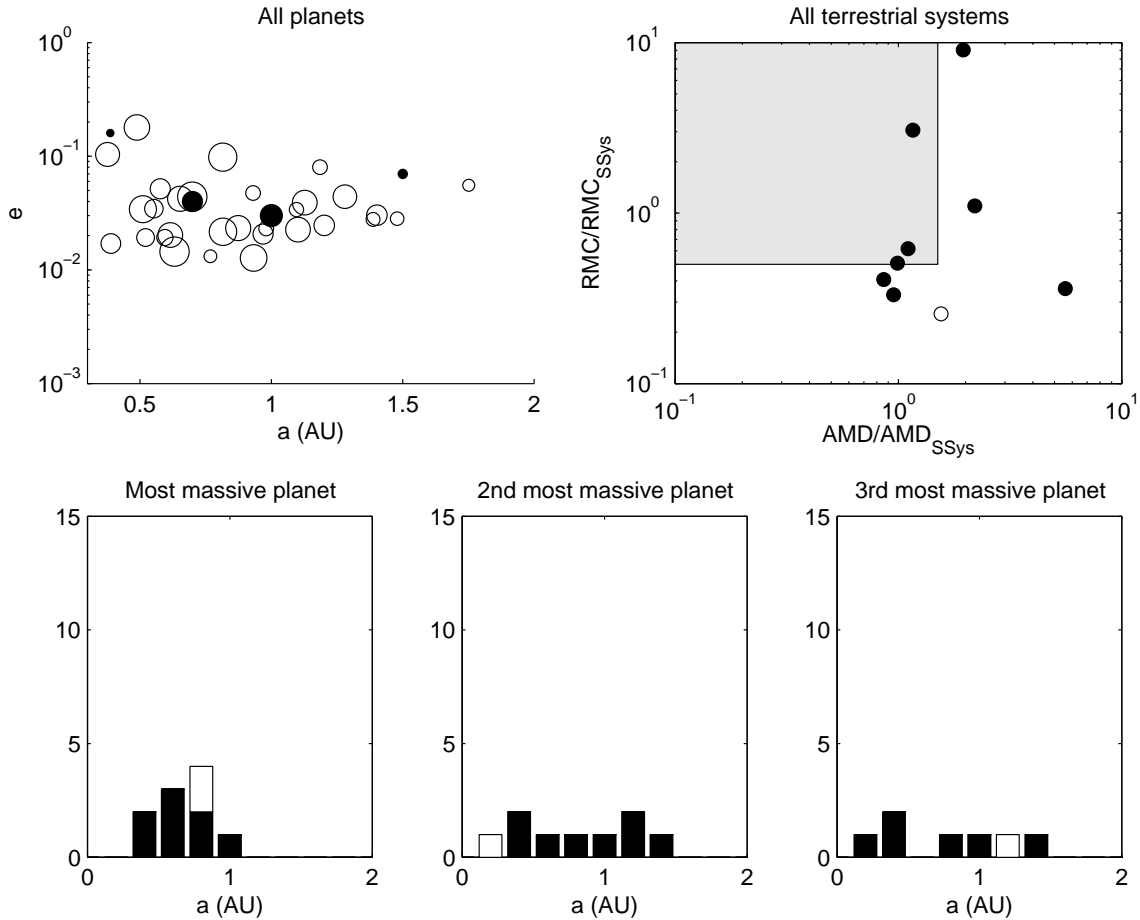} 
\caption{All 10 runs in Set A3 (one system with only a single planet remaining is omitted from the plot).  The only change with respect to Set A1 (Fig. \ref{Set_A1_multipanel}) is that the gas disk height is scaled
up by a factor of 6/5, thus weakening eccentricity damping.}
\label{Set_A3_multipanel}
\end{figure}

Set A3 consists of 10 simulations in which the disk scale height
is increased:
\begin{equation}
H = 0.06 \left ( \frac{r}{\rm 1\,AU}\right )^{5/4}{\rm\,AU}
\label{H=0.06}
\end{equation}
Thus at a given location in the disk, $\tau_{\rm e, tide}$ is
increased by a factor of $\sim 2.4$ relative to Set A1.  The outcomes after, as before, $5 \times 10^8$ years are plotted in Fig. \ref{Set_A3_multipanel}.  Two to three  Solar System analogs are produced, a fraction comparable to the outcome of Set A1 above.  We can see that, in contrast to Set A1, there is an absence of terrestrial systems with an AMD significantly lower than that of the Solar System; in other words, as one might expect, weaker eccentricity damping tends to leave planets more eccentric in the end.  Nevertheless, it appears that, all other
things being equal, decreasing the strength of tidal damping by a
factor of more than two does not have a strong negative effect on
the frequency with which a Solar System-like terrestrial system is
produced by secular resonance sweeping.

\begin{figure}
\plotone{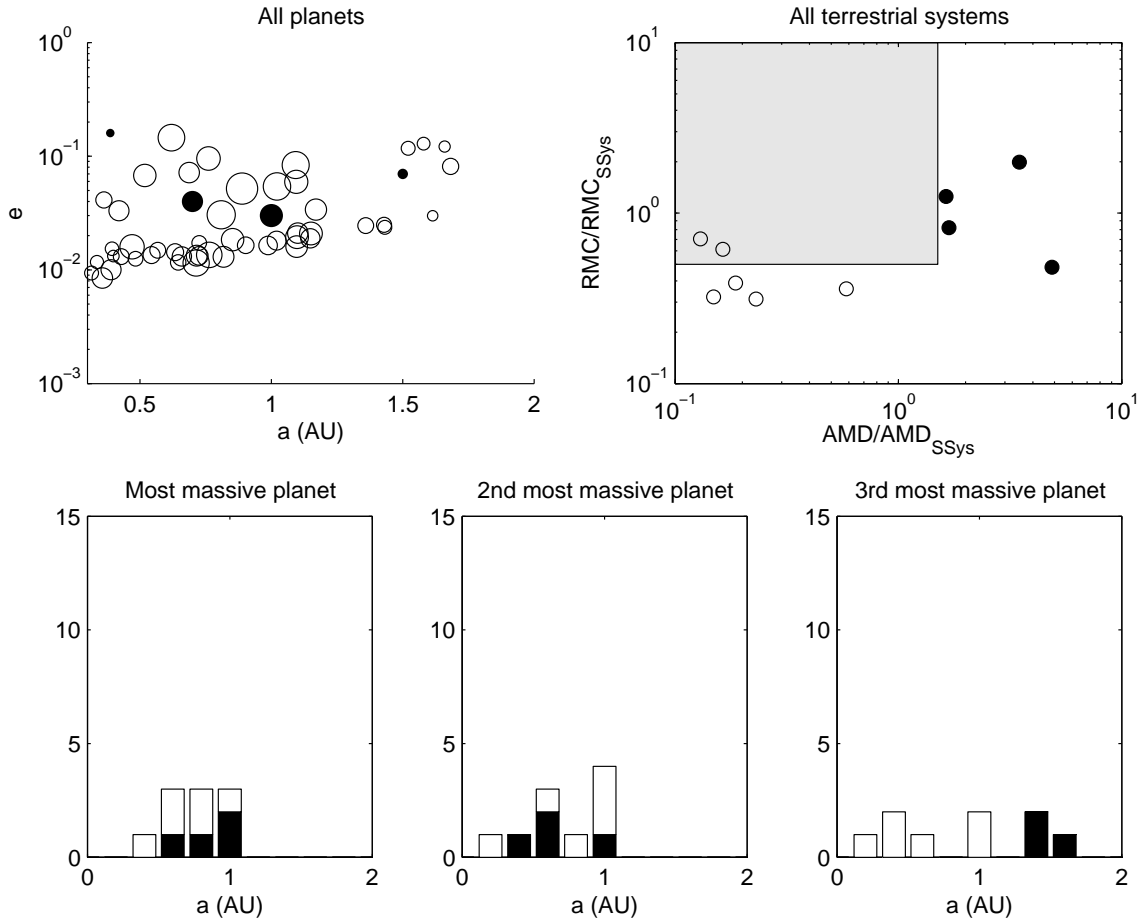} 
\caption{All 10 runs in Set A2.  The only change with respect to Set A1 (Fig. \ref{Set_A1_multipanel}) is that the gas disk height is scaled
down by a factor of 4/5, thus strengthening eccentricity damping.}
\label{Set_A2_multipanel}
\end{figure}

Next, in set A2, we decrease the disk height to
\begin{equation}
H = 0.04 \left ( \frac{r}{\rm 1\,AU}\right )^{5/4}{\rm\,AU}
\label{H=0.04}
\end{equation}
The outcomes are shown in Fig. \ref{Set_A2_multipanel}.  This time, {\it
none} of the resulting systems are Solar System analogs according to our definition.
These
simulations do more poorly because strengthening tidal damping
decreases the secular resonance capture efficiency 
\citep{2005ApJ...635..578N}
As a result, the sweeping $\nu_5$ now leaves
behind a larger number of smaller bodies, rather than retaining
them longer and letting them grow to larger mass as in Set A1 and
A3 above.  We can see that this causes a cluster of low-AMD outcomes with the wrong number of planets (too many).  A second cluster of high-AMD outcomes consists of systems which are left too crowded after the secular resonance sweeps through, and subsequently undergo scattering, producing high eccentricities which can no longer be damped.  

\subsection{A shorter disk depletion time}
\label{set B}

\begin{figure}
\plotone{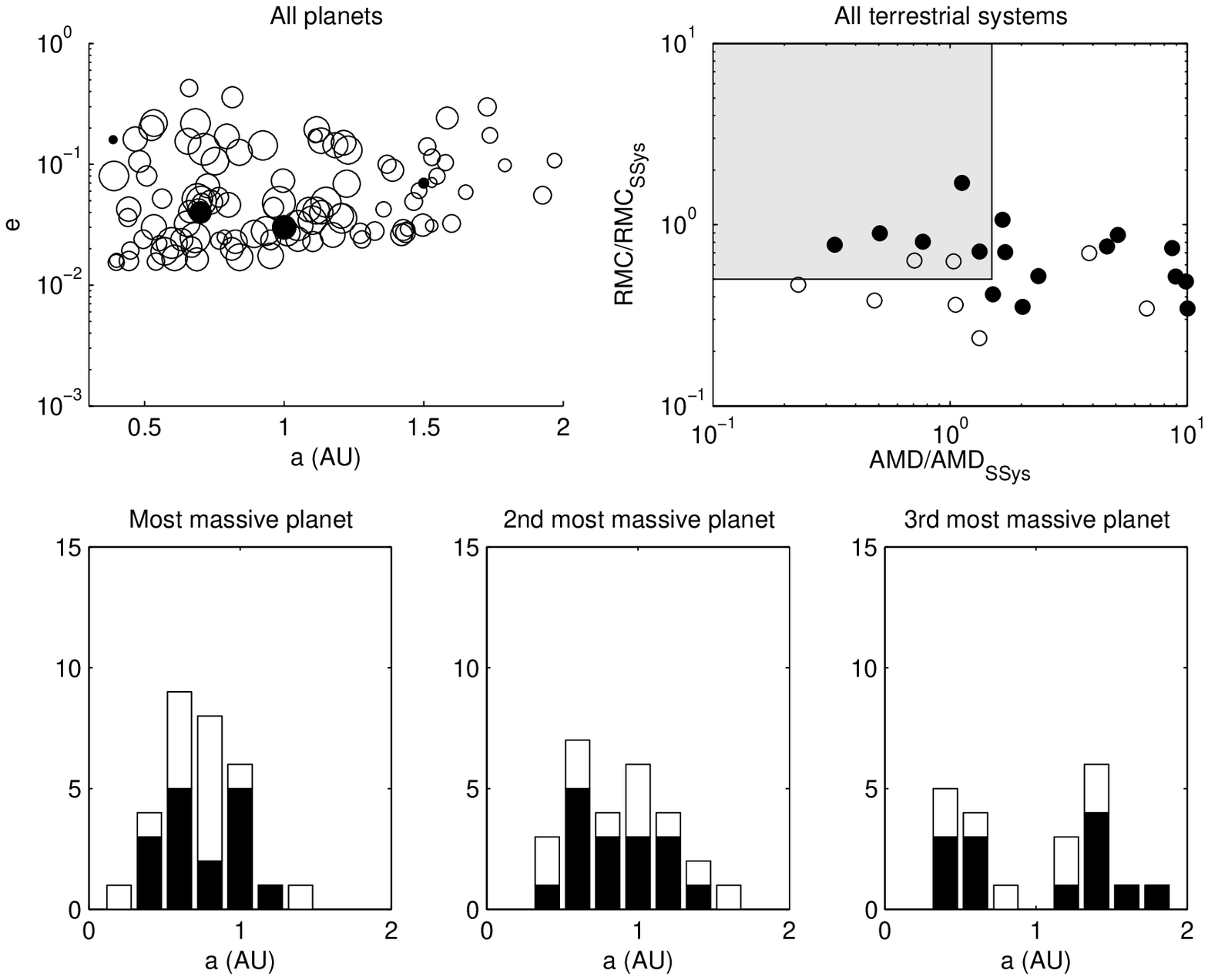} 
\caption{All 30 runs in Set B, encompassing the three different gas disk scale heights.  The only change with respect to Set A1 (Fig. \ref{Set_A1_multipanel}) is that the gas disk depletion timescale is reduced from 5 Myrs to 3 Myrs.}
\label{Set_B_multipanel}
\end{figure}

Observations of T Tauri stars 
(e.g. \citealt{2001ApJ...553L.153H}) show
evidence that disk depletion times can be very short. Accordingly, we perform another set of simulations in which the disk removal timescale $\tau_n$ of Eq. \ref{gas disk Sigma} is shortened from 5 Myrs to 3 Myrs, denoting them Set B.  Since realistic gas disk parameters are rather uncertain and are likely to vary considerably from system to system, we use all three of the disk scale heights above ($H$(1 AU) = 0.06 AU, 0.05 AU, 0.04 AU), performing ten simulations with each and combining the results.  These are shown in Fig. \ref{Set_B_multipanel}.  A total of five Solar System analogs---i.e. $17 \%$ of the total---are produced (as before, none of the small scale height runs contribute).  In comparison, the ``success rate" averaged over Sets A1, A2 and A3 (weighting each set equally)
is $21 \%$.  Thus, the shorter disk depletion time produces a comparable fraction of Solar System-like outcomes.  The planet formation timescale, tied as it is to the disk depletion timescale, is reduced overall by the same factor as the disk depletion time.  Not surprisingly, however, this can only be pushed so far:  Further exploratory simulations---not presented here---suggest that lowering the depletion timescale to $\sim$1 Myr makes Solar System analogs substantially less likely.  When the disk disappears this fast, there is simply not enough time for many mergers to take place before the gas is gone.  Therefore most of the formation process ends up happening in the absence of gas, making the final result very similar to the ``standard model".

\subsection{The effect of lower giant planet eccentricity}
\label{set C}
\begin{figure}
\plotone{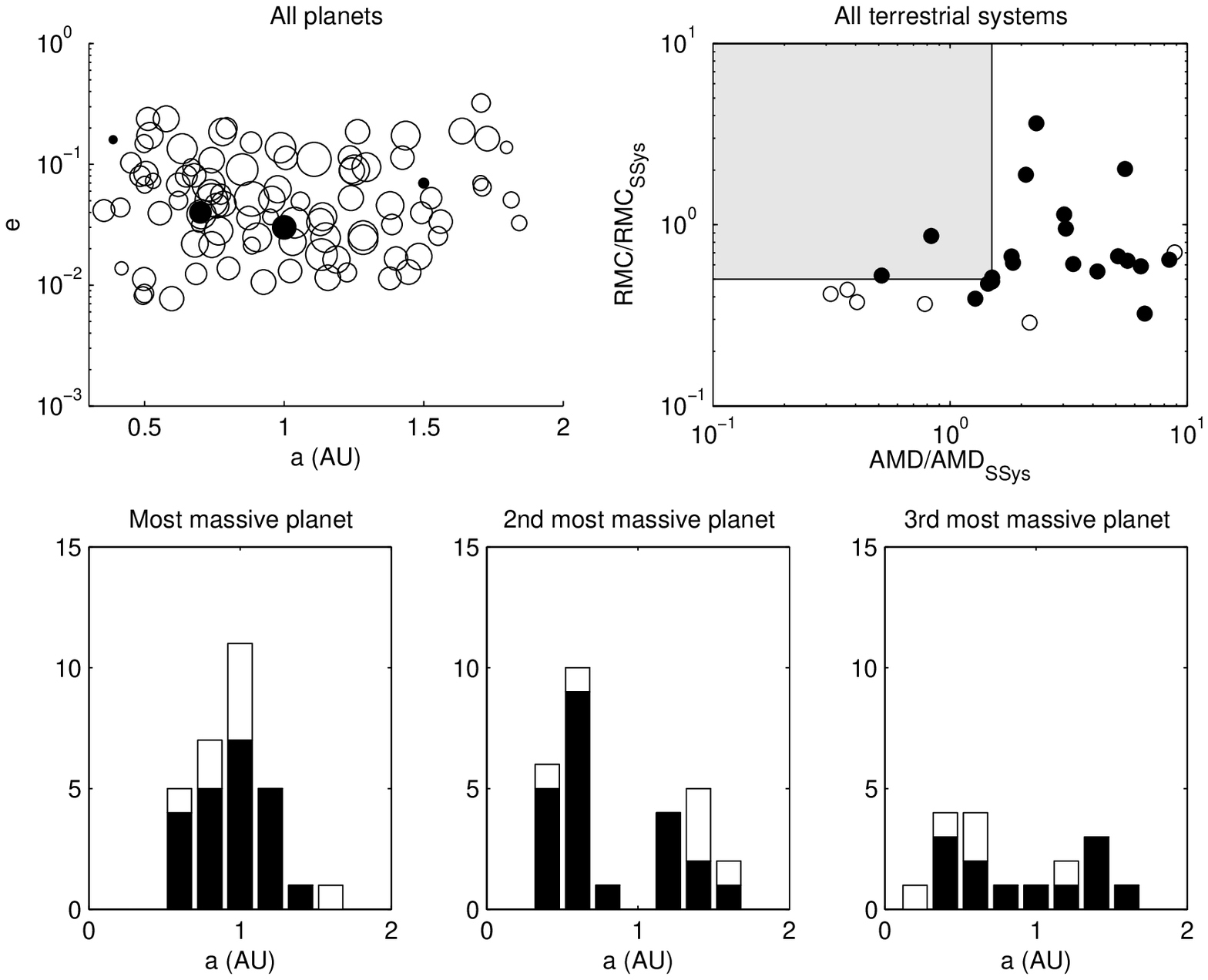} 
\caption{All 30 runs in Set C, encompassing the three different gas disk scale heights.  The only change with respect to
Set A1 (Fig. \ref{baseline_all}) is that the initial giant planet eccentricities are reduced from 0.07 to 0.05.}
\label{Set_C_multipanel}
\end{figure}

We perform another set of 30 simulations, again evenly divided between the three disk thicknesses as in Set B.  The disk depletion time is set back to its baseline value of $5 \times 10^6$ years, but the initial eccentricities of Jupiter and Saturn are reduced from $0.075$ to $0.05$.  Results are shown in Fig. \ref{Set_C_multipanel}.  Only two systems lie within our ``Solar System-like" range.  Most systems have too high an $AMD$.  The reason for the difference is that lower giant planet eccentricities weaken the effect of their secular resonances.  In particular, from Eq. \ref{secular resonance condition} we see that, all other things being equal, reducing Jupiter's eccentricity decreases how far inward a protoplanet is carried before exiting the $\nu_5$ resonance.   The result is a less efficient sweeping process which tends to leave behind many small bodies rather than a few large ones.  
The systems then either remain like this for the full $5 \times 10^8$ yrs of simulation time (the systems below the Solar System range in Fig. \ref{Set_C_multipanel}) or, as a result of their close spacing, undergo interactions at later times which produce more mergers but leave behind high eccentricities and inclinations (systems to the right of the Solar System range).  

Now of course many extrasolar planets have much {\it larger} eccentricities than Jupiter and Saturn, with correspondingly stronger secular resonances.  In such systems, one would reasonably expect secular resonance sweeping to have a greater effect in both clearing planetesimals and promoting planet growth; a detailed investigation will be presented elsewhere.  

\section{Water delivery}
\label{water}
\begin{figure}
\plotone{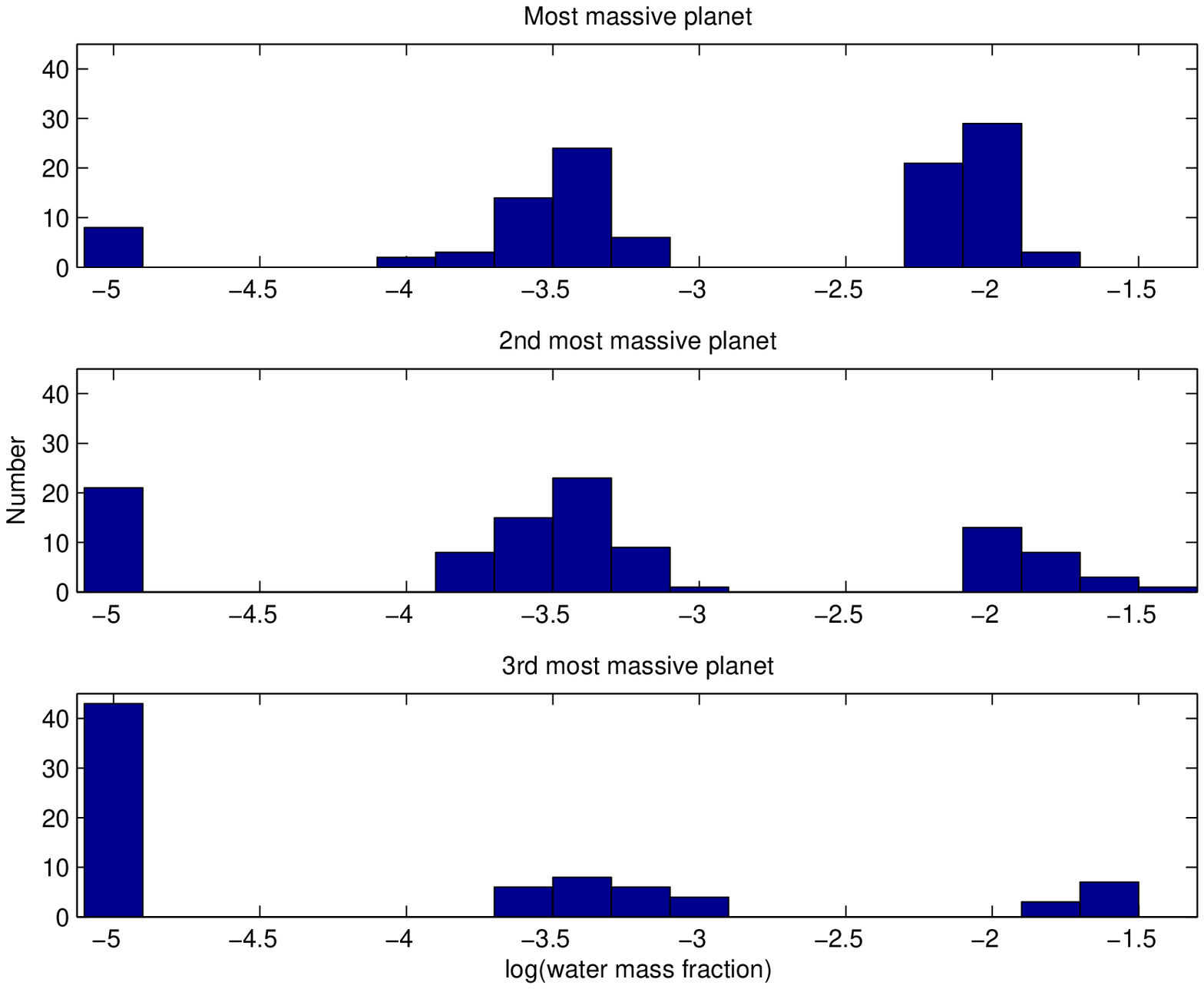}
\caption{Water mass fraction distributions for the most massive, second most massive and third most massive bodies, summed over all 110 simulations presented in this paper (Sets A1, A2, A3, B and C).  The clustering of values occurs because the largest (outermost) initial protoplanets have the highest water content (see text) }
\label{water_histogram}
\end{figure}

Because of the inward delivery of protoplanets by the sweeping secular resonance, material originating well out in what is today the Asteroid Belt region can readily be incorporated into the terrestrial planets.  This has important consequences for the water content of the final products.  As a simple illustration, we use the radial water distribution of 
\cite{2004Icar..168....1R}, with a ``snow line" between 2 and 2.5 AU:
\begin{eqnarray}
M_{\rm water}/M & = & 10^{-5}, r < 2 {\rm AU} \\
\, & = &10^{-3}, 2 {\rm AU} < r < 2.5 {\rm AU} \nonumber \\
\, & = & 5 \times 10^{-2}, r > 2.5 {\rm AU} \nonumber \\
\end{eqnarray}
Using this to assign water mass fractions to our original protoplanets, we can track the accreted material and compute the water mass fractions of the final planets.  Though some water would certainly be lost in collisions, we neglect this effect for simplicity.  Fig. \ref{water_histogram} shows the result for the largest, second largest and third largest planets, summed over all the 110 simulations we performed.  There is a clear trend for larger planets to have a higher water mass fraction.   A part of the reason for this is the outside-in nature of the secular resonance-driven accretion:  Wetter material is captured first, spends the longest time moving with the $\nu_5$ resonance on average, and therefore has the longest window of opportunity for participating in the accretion process.  Most importantly, though, the wettest protoplanets are also initially the most massive (since isolation mass increases with heliocentric distance in our disk model, as is the case whenever the surface density profile is less steep than $\Sigma \propto r^{-2}$).  The initial protoplanet mass at 2 AU is just under 0.2 M$_\oplus$.  The delivery of water-rich material in such large mass increments also accounts for the clustered appearance of the water content in Fig. \ref{water_histogram}.
\section{Discussion}
\label{discussion}
The simulations we perform here illustrate a scenario for the
final stage of forming the Solar System terrestrial planets which
differs fundamentally from the ``standard model".  Rather than the
isolation-mass bodies produced by oligarchic growth gradually
perturbing each other onto crossing orbits, the whole process is
driven by the depletion of the gas disk.  As Jupiter's $\nu_5$
secular resonance sweeps inward, the eccentricity excitation it
produces, combined with tidal eccentricity damping by the gas
disk, induces protoplanets to migrate in with the resonance.
Neighboring protoplanets are thus brought into close proximity.   
As discussed in 
\cite{2005ApJ...635..578N}, even though significant eccentricities are produced by the secular
resonance, a tendency toward local apsidal alignment of
protoplanet orbits keeps relative velocities low, enhancing the effectiveness of gravitational focusing and thus increasing the rate at which they accrete each other.  In the
idealized picture, then, protoplanets migrate inward with the
sweeping $\nu_5$, accreting each other along the way, until Mars,
Earth and then Venus are formed and left behind.  With a small
fraction of the nebular gas still present, these planets'
eccentricities get damped to their current low values.  The relatively high mass concentration within the Solar System's terrestrial region, centered near 1 AU, results because the $\nu_5$ resonance preferentially deposits Earth/Venus-mass planets in this region (Eq. \ref{secular resonance condition}). 

In practice of course, even though the sweeping secular resonance imposes some order, and a
definite timescale, on the last stage of terrestrial planet
formation, what results is still a highly stochastic process.  Thus the successful formation of a good analog to the terrestrial region of the Solar System only takes place a fraction of the time.  For the baseline version of our model (sets A1, A2 and A3, \S \ref{baseline_case} and \S \ref{changing t_e}, the ``success rate" is 21 \%.  In fact, 9\% are {\it highly} successful, in the sense that they posses both a lower angular momentum deficit and a higher radial mass concentration than the Solar System (see Fig. \ref{Set_A1_multipanel}).  Shortening the e-folding time for disk dispersal from 5 to 3 Myrs (set B, \S \ref{set B}) yields a similar success rate of 17\%.  Initial indications are that further reducing the disk dispersal timescale to 1 Myr results in close to zero success rate; more work is need to investigate this.  Reducing the initial eccentricity of Jupiter from 0.075 to its current value of 0.05 (set C, \S \ref{set C}) also drops the success rate to near zero.  Both these parameters influence the effectiveness of the sweeping secular resonance in capturing and retaining protoplanets, which is of central importance to the functioning of the dynamical shakeup model.  The latter result suggests the need for a slightly elevated primordial Jovian eccentricity, as might be produced either by close encounters among growing proto-gas giants, or by planet-disk interaction once Jupiter opens its gap; the same mechanisms, in other words, which are also invoked  as candidates for producing the substantial eccentricities observed in exoplanets, though in this case with a relatively mild outcome.  Some of the eccentricity is damped by the indirect feedback of the gas disk on Jupiter---through interaction with the damped protoplanets---while additional damping would occur later as Jupiter scatters remnant planetesimals from the outer Solar System.  Also, the initial mass distribution in our simulations has an outer edge at 3 AU; including more of the material in the outer belt region further adds to the damping on Jupiter (\citealt{2006Icar..184...39O} and O'Brien, private communication).

The latter process is not modeled, thus final giant planet eccentricities are still $\sim 0.07$ for sets A and B.  For set C, they are $\sim 0.03-0.04$.  Conversely, if planet-disk interaction in fact damped eccentricities to values {\it lower} than their current values, then secular resonance sweeping would have been unlikely to work.   

We thus arrive at the conclusion that this model provides a plausible pathway to producing an analog to the Solar System's terrestrial planets.  In contrast, the gas-free standard model of terrestrial planet formation does not appear to provide any pathway, since the above two key parameters---angular momentum deficit ($AMD$) and radial mass concentration ($RMC$)---are {\it never} reproduced.  The simulations of \cite{2001Icar..152..205C}, for example, have a minimum $AMD$ of 1.8 times that of the Solar System, and a maximum $RMC$ of 0.71 times that of the Solar System.  It should be pointed out that the situation changes if the standard model is modified with the assumption that the  Solar System retains a substantial population of planetesimals into the giant-impact phase of formation. In this case, it is possible to produce a good match to the Solar System's $AMD$ \citep{2006Icar..184...39O} ($RMC$ values remain comparable to those of \cite {2001Icar..152..205C}; O'Brien, private communication).  However, such persistent planetesimals raise both theoretical and observational concerns (\S \ref{intro}).

At the same time, however, the majority of the outcomes are ``failures" in the sense that they do not resemble the Solar System.  They are nevertheless interesting, since they may be representative of the outcomes of terrestrial planet formation in other planetary systems.  Generally speaking, failures result when terrestrial planet formation has not finished (or nearly finished) after a few disk dispersal e-folding times.  Then, the outcomes can be grouped into two broad categories: (i) systems which have high eccentricities, because the formation process finished in the absence of gas, and (ii) systems which remain with larger numbers of bodies, yet are stable at least on the 0.5 Gyr time of our simulations.  The former cases are not distinguishable from the ``standard model" where all of late-stage accretion plays out in the absence of gas.  The latter, however, are interesting.  These precarious-looking yet stable configurations arise due to the damping effect of the disk, which in the end leaves the eccentricities low enough for long-term stability.  Such crowded terrestrial systems may await discovery around other stars, and raise the astrobiologically intriguing possibility of multiple planets sharing a habitable zone (e.g. panels 5, 6, 16 and 24 of Fig. \ref{baseline_all}).  Overall, the rich variety of our results suggests that diverse evolutionary paths are readily explored by different systems during the transition from protostellar to debris disks. Spitzer data show large dispersions in the IR excess of intermediate-mass young stars with age ranging from a few Myr to a few 
   tens Myrs 
\citep{2005ApJ...620.1010R}.
Sweeping secular resonances by putative 
eccentric planets in depleting disks may be one of the contributing factors to the
   observed dispersion.

Although individual cases are subject to large stochastic variations, there are several features common to most or all of our simulation outcomes.  One is the large net inward migration of mass.  We
extend our starting population of isolation-mass protoplanets to
almost 3 AU, far out into present-day the asteroid belt, yet in all cases the two most massive planets end up inside 1.5 AU.  In fact, very few simulations end with {\it any} planets outside this radius.  This makes it likely that significant amounts of water-rich material
will end up incorporated into the final planetary system (\S \ref{water}).
 Another
relatively common feature is the production of Mars-like planets;
that is, small bodies
which end
up as the outermost of the final planets. These result from close
encounters with other bodies, which scatter them out of the secular
resonance. Early on, when there are many bodies moving with the
$\nu_5$, the smallest ones---often just single unaccreted
oligarchs---are at the greatest risk of being scattered and left
behind.  Later on, when there are fewer bodies and the resonance
gets weaker, smaller bodies have the advantage because they are
better able to remain entrained in the resonance (Eq. \ref{secular resonance condition}).  In contrast, Mars analogs tend to be  rare in numerical simulations of the standard model.  Finally, a feature common to all simulations is
simply that 
collisional evolution proceeds rapidly; an initial population of 40 or
more protoplanets is reduced to usually well below ten merger
products in $\sim$ 4 times the disk depletion time ($4 \times 5=20$ Myrs in Sets A and C, $4 \times 3=12$ Myrs in Set B).  

In \S \ref{baseline_case}, we argued that the apparent lack of substantial migration by Jupiter and Saturn in the early Solar System
makes it reasonable to start the two planets with their current semimajor axes.  However, if significant planetesimal-driven migration occurs after all the gas has been removed, then the last part of the $\nu_5$'s journey to its present location will take place after terrestrial planet formation is complete.  Further simulations would be required to assess the effect on the formation process, though one can anticipate that if the resonance remains well beyond 1 AU during this time, terrestrial planets will also tend to be left at larger radii.   The scenario of  
\cite{2005Natur.435..466G}, in which all the giant planets start out in a very compact configuration and remain that way until the Late Heavy Bombardment, when the Solar System has an age of $\sim 700$ Myrs, would constitute an extreme version of this.  In fact, such late migration poses a worrisome problem in {\it any} scenario of terrestrial planet formation:  Since the $\nu_5$ resonance is currently just inside Venus' orbit, late divergent migration of Jupiter and Saturn would have to involve Venus being crossed by the resonance.  Because this happens in the absence of damping, it is difficult to understand how Venus could maintain its low eccentricity in this scenario.

Parent-daughter pairs of radioactive isotopes can be exploited as chronometers to trace the timelines of processes in the early Solar System, potentially giving us additional constraints on formation models.  Two such pairs are Hf-W and U-Pb.  For both of these, the parent element is a lithophile that is retained in silicate reservoirs during planetary accretion, whereas the daughter element is segregated into the core.  Comparisons of the Hf-W ratio in the Earth's mantle 
to those of meteorites have been used to derive a growth time of the Earth ranging from about 11 to 50 Myrs \citep{2002Natur.418..949Y,2002Natur.418..952K,wood_halliday_05}.  In particular, the derived timescale is $\ga 30$ Myrs
if the Earth's accretion was terminated by the Moon-forming impact \citep{2001Natur.412..708C}, assuming this produced complete metal-silicate equilibration.  This result is intriguingly similar to the short timescales for terrestrial planet formation we obtain here.
However, the U-Pb chronometer implies a later formation time, $\sim 65-85$ Myrs
\citep{2004Natur.431..253H}.
It may be that U-Pb traces the last stages of core segregation, while Hf-W traces the time for {\it most} of the core to finish forming 
\citep{2005prpl.conf.8221S}.
Thus the picture from cosmochemistry is not yet conclusive, but chances are good that future developments will tell us with some certainty whether the formation of the terrestrial planets was rapid or drawn-out.  

\acknowledgements
We are grateful to the referee, David P. O'Brien, for a thorough report which helped us to improve the paper.  EWT acknowledges support by the NSF through grant AST-0507727 at Northwestern University, by CITA, and by Canada's National Sciences and Engineering Research Council.  MN acknowledges support by MEXT (MEXT's program ``Promotion of Environmental Improvement for Independence of Young Researchers" under the Special
Coordination Funds for Promoting Science and Technology and
``KAKENHI-18740281" Grant-in-Aid for Young Scientists B).  DNCL acknowledges support by NASA (NAGS5-11779, NNG04G-191G,
NNG06-GH45G, NNX07AL13G, HST-AR-11267), JPL (1270927), and the NSF(AST-0507424).


\end{document}